\documentclass{PoS}

\title{How warm is the molecular gas in active environments?}

\ShortTitle{How warm is the molecular gas in active environments?}

\author{\speaker{Stefanie M\"uhle}\\     
        Joint Institute for VLBI in Europe \\
        Postbus 2, 7990 AA Dwingeloo, The Netherlands\\
        E-mail: \email{muehle@jive.nl}}

\author{Christian Henkel\\
        Max-Plank-Institut f\"ur Radioastronomie\\
        Auf dem H\"ugel 69, 53121 Bonn, Germany\\ 
        E-mail: \email{p220hen@mpifr-bonn.mpg.de}}

\author{Tahlia de Maio\\
        University of Colorado\\
        389 UCB, Boulder CO, 80309, USA\\
        E-mail: \email{tahlia.demaio@colorado.edu}}

\author{Ernest R.\ Seaquist\\
        Department of Astronomy and Astrophysics, University of Toronto\\
        50 St. George Street, Toronto, ON M5S 3H4, Canada\\
        E-mail: \email{seaquist@astro.utoronto.ca}}

\abstract{The question whether or not the initial mass function is universal, 
i.e.\ the same in all kinds of environments, is of critical importance for the 
theory of star formation and still intensely debated.
A top-heavy initial mass function may be the result of star formation 
out of dense molecular clouds with a temperature of $\sim 100$\,K. Such a 
molecular gas phase is not commonly found in the Galactic plane, but may be 
present in active environments like cores of starburst galaxies or AGN. 
Unfortunately, the kinetic temperature of the molecular gas in external 
galaxies is often not well constrained. Having proven the diagnostic power 
of selected formaldehyde lines as tracers of the properties of the molecular 
gas in external galaxies, we have engaged in observing these diagnostic lines 
in a number of starburst galaxies or near AGN. This contribution presents the 
latest results of these studies.}

\FullConference{10th European VLBI Network Symposium and EVN Users Meeting: 
                VLBI and the new generation of radio arrays \\
                September 20-24, 2010\\
                Manchester, UK}

\newcommand{\htco}{H$_2$CO}
\newcommand{\kms}{km\,s$^{-1}$}

\begin{document}

\begin{figure}[h]
\begin{center}
\includegraphics[width=.6\textwidth]{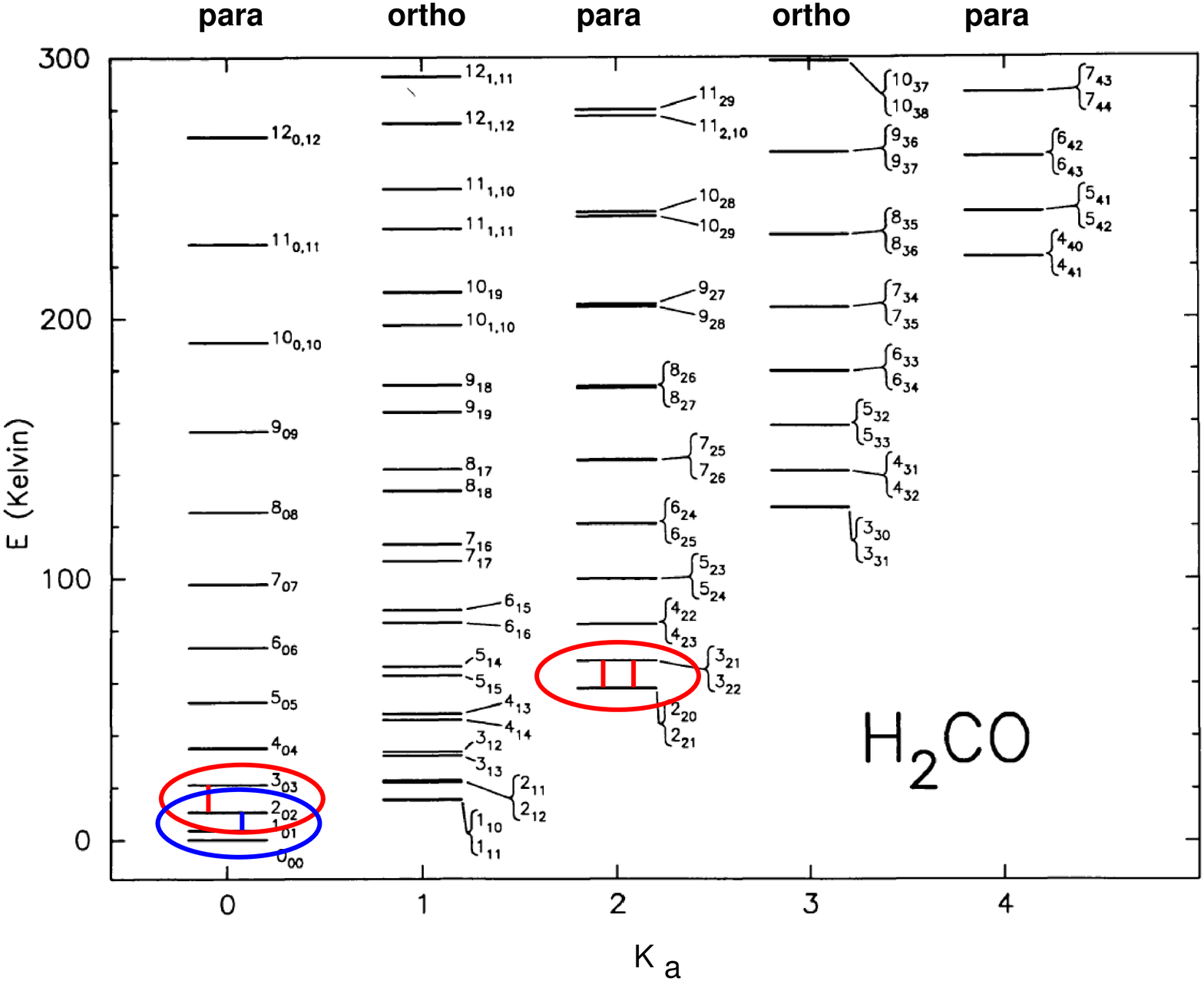}
\end{center}
\caption{The lowest energy levels of ortho- and paraformaldehyde, adapted from 
         [7]. The para-H$_2$CO transitions at 218\,GHz are marked in red, the 
         diagnostic para-H$_2$CO transition at 146\,GHz in blue.}
\label{fig1}
\end{figure}

\section{Is the initial mass function universal?}

One of the fundamental debates in the field of star formation is the question  
whether the stellar initial mass function (IMF) 
is universal or not. A number of recent observations have been interpreted as 
evidence for a top-heavy initial mass function, spanning a variety of objects, 
from the centre of our Galaxy to circumnuclear starburst regions and 
ultra-compact dwarf galaxies (e.g.\ [3,4], see also [1] for a critical review). 
Hydrodynamical simulations can reproduce such a top-heavy IMF if the the raw 
material of star formation, the dense molecular gas, is assumed to have a 
kinetic temperature of $\sim 100$\,K [6]. 
Such a temperature is significantly higher than what is observed in the dense 
cores in the Galactic plane, but similar to kinetic temperatures that 
have been derived for a number of starburst galaxies and AGN from observations 
of highly excited ammonia lines and IR rotational H$_2$ lines [10,14]. 
The standard conversion factor between the CO(1-0) line intensity and the 
H$_2$ column density that has been derived empirically in the Galactic plane 
can be up to an order of magnitude too large in starburst galaxies. This 
supports the view that the properties of the molecular gas in active 
environments like the cores of starburst galaxies or AGN, whose activity leads 
to strong feedback in the form of strong UV radiation fields and high 
intensities of X-rays and cosmic rays, can differ significantly from the 
properties of the molecular gas in the Galactic disk, in particular in 
the dense cores [13].

\begin{figure}[h]
\includegraphics[width=.9\textwidth]{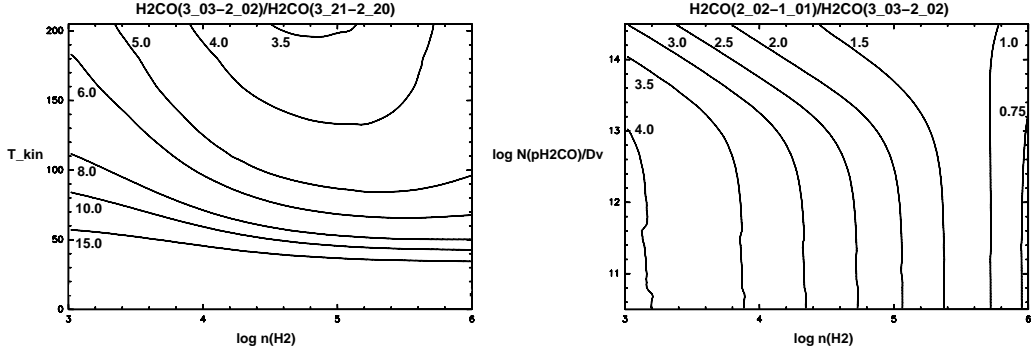}
\caption{Plots derived from our LVG code illustrating the diagnostic value of
the selected line ratios: {\bf a)} The interladder line ratio
H$_2$CO($3_{03} \to 2_{02}$)/H$_2$CO($3_{21} \to 2_{20}$) traces the kinetic
temperature with only a weak dependence on the gas density or the para-H$_2$CO
column density (here: $N_{\rm pH2CO}/\Delta v =10^{12}\,{\rm cm}^{-2}/{\rm km\,s^{-1}}$).
{\bf b)} At a given kinetic temperature (here: $T_{\rm kin}=100$\,K), the
intraladder line ratio H$_2$CO($2_{02} \to 1_{01}$)/H$_2$CO($3_{03} \to 2_{02}$)
provides an excellent estimate of the gas density with only a moderate
dependence on the para-H$_2$CO column density.}
\label{fig2}
\end{figure}

\section{Formaldehyde lines as extragalactic diagnostics}

Unfortunately, the kinetic temperature of the molecular gas in most external 
galaxies is not well constrained yet, because many of the most common tracer 
molecules suffer from a degeneracy between the kinetic temperature and the gas 
density.
Instead, the kinetic temperature of the molecular gas is often set to be 
similar to the dust temperature, an assumption that may be valid in the 
densest, most FUV-shielded cores of giant molecular clouds, but can only be a 
lower limit in most other cases. Ammonia, a molecule that thanks to its 
symmetric top structure is well suited as a temperature probe, serves 
well as the standard ``cloud thermometer'' in our Galaxy where the difference 
between the derived rotational temperature and the kinetic temperature is 
small. Unfortunately, the fractional abundance of ammonia varies between 
$10^{-5}$ in hot cores [9] and $10^{-8}$ in dark clouds [2], which makes this 
molecule less suitable as a tracer in extragalactic observations that by 
default average over a significant part of a galaxy.

In contrast, formaldehyde (H$_2$CO), an only slightly asymmetric top, shows 
little variation in its fractional abundance in a variety of galactic 
environments, with infrared sources showing the largest discrepancy with a 
factor of $5-10$ [5]. The two subspecies para- ($K_a=0$, 
2, 4, $\ldots$) and orthoformaldehyde ($K_a=1$, 3, 5, $\ldots$) possess a rich 
spectrum with a number of transition lines in the cm- and mm-range that can be 
observed with existing telescopes (Fig.~1). On the other hand, these radiative 
transitions are not so numerous that line blending, which is commonly found in 
complex molecules, is a major concern. As a rule of thumb, line intensity 
ratios that involve different $K_a$ ladders (interladder ratios) are good 
tracers of the kinetic temperature, while intraladder ratios, i.e. ratios of 
transitions of the same $K_a$ ladder, sensitively probe the gas density once 
the temperature is determined (Fig.~2, [7]). 

\begin{figure}[h]
\includegraphics[width=.9\textwidth]{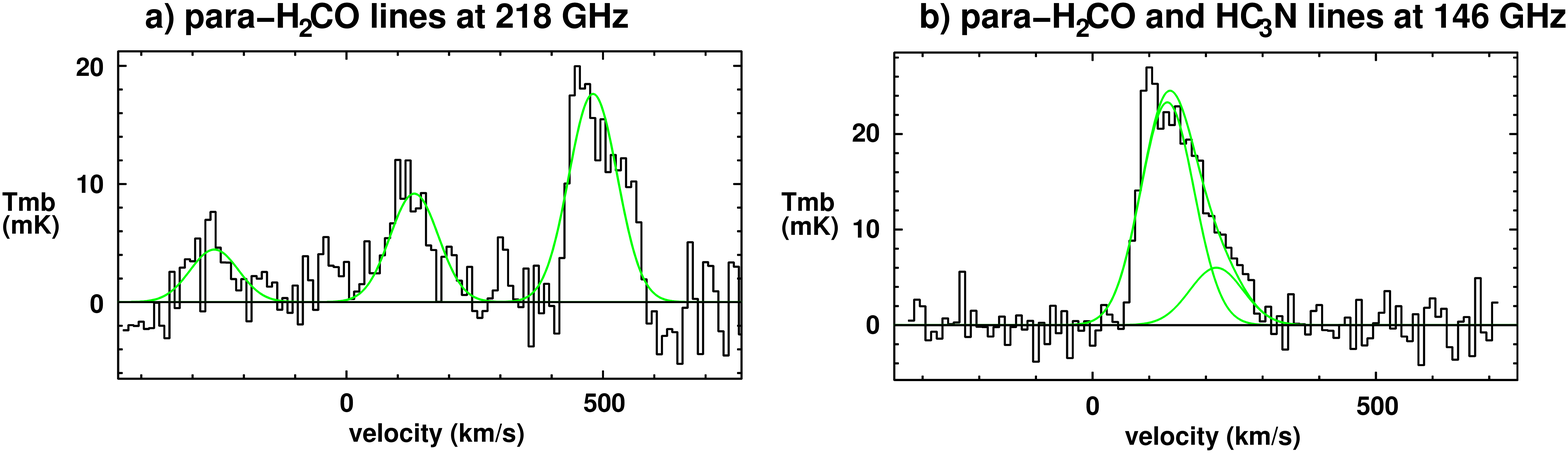}
\caption{Paraformaldehyde lines observed towards the SW lobe of M\,82. 
{\bf a)} Para-H$_2$CO lines at 218\,GHz. The velocity scale of the spectrum 
refers to the \htco($3_{22} \to 2_{21}$) line at 132\,\kms . The 
\htco($3_{03} \to 2_{02}$) and the \htco($3_{21} \to 2_{20}$) transitions are 
offset by 348.5\,\kms\ and $-389.7$\,\kms, respectively. 
{\bf b)} In the 146\,GHz spectrum, the \htco($2_{02} \to 1_{01}$) line at 
132\,\kms is blended with the HC$_3$N($16 \to 15$) line, which is offset by  
86.46\,\kms\ from the \htco($2_{02} \to 1_{01}$) transition.
The green curves show the Gaussian fit to each individual line as well as the superposition of all fitted Gaussians.}
\label{fig3}
\end{figure}

\section{The project}

Considering the important role of the kinetic temperature of the star-forming 
molecular gas in active environments on one hand and the scarcity of 
reliable temperature measurements in external galaxies on the other hand, 
we have engaged in deriving the properties of the dominant phase of the 
molecular gas in a number of nearby starburst galaxies and AGN using the 
diagnostic properties of paraformaldehyde lines at 146\,GHz and at 218\,GHz. 
The paraformaldehyde lines at 218\,GHz are particularly valuable from a 
diagnostic point of view. Within 1\,GHz, i.e. usually within the same low 
frequency resolution spectrum, there are the three para-H$_2$CO transitions 
$3_{03} \to 2_{02}$ (218.22\,GHz), $3_{22} \to 2_{21}$ (218.48\,GHz) and  
$3_{21} \to 2_{20}$ (218.76\,GHz) (Fig.~3a). While the 
H$_2$CO($3_{22} \to 2_{21}$) line may be contaminated by methanol emission, the 
intensity ratio of the other two lines forms an excellent temperature tracer, 
without any uncertainty due to calibration issues, pointing errors or 
different beam widths. The set of diagnostic lines is completed by 
H$_2$CO($2_{02} \to 1_{01}$) at 145.60\,GHz, an easily detectable line that 
together with H$_2$CO($3_{03} \to 2_{02}$) forms a sensitive density tracer.\\
The reasonable assumption that the emission of all the H$_2$CO lines 
originates from the same volume implies that the lines have the same velocity 
profile, thus considerably reducing the number of free parameters in the 
derivation of the line intensities: If the line profiles are approximated by 
Gaussian curves, the velocity and width of all lines should match, leaving the
peak intensity as the only free parameter. Thus, by fixing the velocity and 
the width of the common line profile with the help of a strong line, even the
intensities of weak (Fig.~3a) or blended lines (Fig.~3b) can be derived with 
high reliability.

By comparing the line intensity ratios with the predicted ratios of a non-LTE 
model physical properties of the gas like the kinetic temperature and the 
average gas density within the molecular cloud complexes can be derived. 
For the analysis of the para-H$_2$CO lines, we have developed a model with a 
spherically symmetric cloud geometry and adopted the LVG approximation. The 
searched parameter space covers a kinetic temperature of $T_{\rm kin}=5$ to 
300\,K in steps of 5\,K, a molecular gas density of $\log{n_{\rm H2}}=3.0$ to 
6.0 (in cm$^{-3}$) in steps of 0.1 and a para-H$_2$CO column density per 
velocity interval of $\log{N_{\rm pH2CO}/\Delta v}=10.5$ to 14.5 
(in cm$^{-2}$\,km$^{-1}$\,s) in steps of 0.1 (see [11] for details). 

\begin{figure}
\includegraphics[width=.95\textwidth]{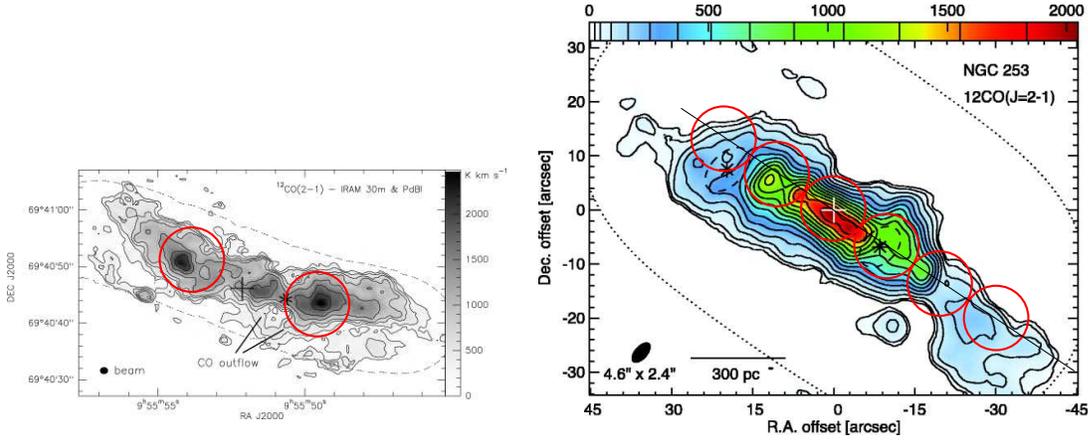}
\caption{Positions of observed \htco\ emission in M\,82 (left) and NGC\,253 (right) superposed on high-resolution maps of the integrated CO($2 \to 1$) intensity adapted from [17] and [15], respectively. The diameter of the red circles corresponds to the width of the beam of the IRAM 30-m telescope at 218\,GHz.}
\label{fig4}
\end{figure}

\section{First results}

The first results of our studies support the view that a considerable fraction 
of the dense molecular gas in starburst galaxies is significantly warmer than 
the dense molecular gas in quiescent galaxies. A comparison between M\,82 and 
NGC\,253 shows interesting similarities and differences. Both galaxies are 
considered to be prototypical starburst galaxies, nearby and seen at high 
inclination, with high infrared luminosities, super-star clusters and a 
prominent galactic wind. In both galaxies, the dense molecular gas is 
concentrated around the starbursting nucleus. The fractional abundances of a 
number of molecules, though, differ strongly between the two galaxies, which 
is seen as evidence that the starburst in M\,82 is in an evolved state while 
the 
starburst in NGC\,253 is (kept) young. Note that the massive star cluster M82F
is to date the best candidate to host a top-heavy IMF [1].

We observed the two lobes of the circumnuclear ring in M\,82 and six positions 
along the major axis of the circumnuclear disk of NGC\,253 with the IRAM 30-m 
telescope both at 218\,GHz and at 146\,GHz to an rms noise level of a few 
milliKelvin at a velocity resolution of 10\,\kms\ at 
each frequency and pointing position (Fig.~4). The \htco($2_{02} \to 1_{01}$) 
line at 145.60\,GHz, slightly blended with the HC$_3$N($16 \to 15$) line, has 
been detected at all pointing positions, while the three \htco\ lines at 
218\,GHz have been deteced at all positions except for the two outermost 
pointings in NGC\,253. As an example, the spectra of the southwestern lobe in 
M\,82 are shown in Fig.~3. 

The intensity ratios derived from the observed lines are remarkably similar in 
the two lobes of M\,82, whereas the ratios vary significantly along the major 
axis of the molecular disk in NGC\,253, indicating different physical 
properties of the dense molecular gas across the galactic core region.
In M\,82, the LVG analysis suggests a kinetic temperature of 
$T_{\rm kin}\sim 200$\,K, a gas density of 
$n_{\rm H2} \sim 7 \cdot 10^3$\,cm$^{-3}$ and a molecular gas mass of 
$\sim 3 \cdot 10^8\,M_{\odot}$ in the lobes, which is in good agreement with 
the high-excitation molecular gas component of other recent molecular line 
studies, which may constitute more than half of the total molecular gas mass 
in M\,82 [8,16].
A preliminary analysis of the line ratios derived for the different positions 
in NGC\,253 yields kinetic temperatures of $T_{\rm kin} \sim 70 \ldots 150$\,K 
and gas densities of $n_{\rm H2} \sim 5 \ldots 25 \cdot 10^3$\,cm$^{-3}$. Thus,
in both galaxies, the dense molecular gas component traced by paraformaldehyde 
lines is considerably warmer than the dust and in fact warm enough to give 
rise to a top-heavy IMF should this gas be turned into stars.

\section{Outlook}

The survey of the temperature of dense molecular gas in active extragalactic 
environments is still in an early phase and many questions about the extent of 
the warm gas phase and the environments where it can be found remain. 
ALMA will provide a boost in sensitivity which will make the para-\htco\ lines 
readily detectable in a variety of external galaxies, while at the same time 
offering unprecedented angular resolution. This will provide information on the 
distribution of the traced molecular gas phase and constrain source sizes. In
the cm-range, e-MERLIN may be used to supply high-resolution maps of the 
ortho-\htco\ $K$-doublet transition $1_{10} \to 1_{11}$, which will allow us to
derive the ortho-to-para formaldehyde ratio, thus providing information
on the temperature of the dust grains at the time when the formaldehyde 
molecules were formed.

\acknowledgments
We wish to thank the staff at Pico Veleta/Granada for their support during 
the observations. E.R.S. acknowledges a Discovery Grant from NSERC.  
This work has made use of the following software and resources: GILDAS, specx, 
Statistiklabor (FU Berlin, CeDiS), asyerr (Seaquist \& Yao), 
the JPL Catalog of spectral lines, The Cologne Database for Molecular 
Spectroscopy [12] and NASA's Astrophysics Data System 
Bibliographic Services (ADS).

\end{document}